%% file: main.tex
\documentclass{sigchi-ext}
\usepackage[T1]{fontenc}
\usepackage{textcomp}
\usepackage[scaled=.92]{helvet} 
\usepackage{graphicx} 
\usepackage{balance}  
\usepackage{booktabs} 
\usepackage{ccicons}  
\usepackage{ragged2e} 

\usepackage{color}
\usepackage[usenames,dvipsnames]{xcolor}
\usepackage{textcomp}
\usepackage{booktabs}
\usepackage{todonotes}
\usepackage{xcolor}
\usepackage{adjustbox}
\usepackage{booktabs}
\usepackage{graphicx}
\usepackage{caption}

\usepackage{hyperref}
\hypersetup{
    colorlinks=true,
    linkcolor=black,
    urlcolor=blue,
}
 
\usepackage{import}

\copyrightinfo{
 Datasets are available online at \url{https://research.arch.tamu.edu/analytic-provenance/datasets/} for research purposes.}
 
\title{Analytic Provenance Datasets: A Data Repository of Human Analysis Activity and Interaction Logs}

\numberofauthors{6}
\author{%
  \alignauthor{%
    \textbf{Sina Mohseni}\\
    \affaddr{Department of Computer Science \& Engineering}\\
    \affaddr{Texas A\&M University}\\
    \affaddr{sina.mohseni@tamu.edu} }
  \alignauthor{
    \textbf{Andrew Pachuilo}\\
    \affaddr{Department of Computer Science \& Engineering}\\
    \affaddr{Texas A\&M University}\\
    \email{apachuilo@tamu.edu} 
    }
     \vfil
  \alignauthor{%
    \textbf{Ehsanul Haque Nirjhar}\\
    \affaddr{Department of Computer Science \& Engineering}\\
    \affaddr{Texas A\&M University}\\
    \affaddr{nirjhar71@tamu.edu} }
  \alignauthor{%
    \textbf{Rhema Linder}\\
    \affaddr{Department of Computer Science \& Engineering}\\
    \affaddr{Texas A\&M University}\\
    \affaddr{rhema@tamu.edu} 
    }
     \vfil
  \alignauthor{
    \textbf{Alyssa~Pe\~{n}a}\\
    \affaddr{Department of Visualization}\\
    \affaddr{Texas A\&M University}\\
    \email{mupena17@tamu.edu} } 
  \alignauthor{
    \textbf{Eric D. Ragan}\\
    \affaddr{Department of Visualization}\\
    \affaddr{Department of Computer Science \& Engineering}\\
    \email{eragan@tamu.edu} \\
} }

\def\plaintitle{Analytic Provenance Datasets: A Data Repository of Human Analysis Activity and Interaction Logs} \def\plainauthor{Sina Mohseni, Eric D. Ragan}
\def\plainkeywords{Analytic provenance; Text analysis; Cyber analysis; User interaction logs; Eye tracking; Online dataset.}

\hypersetup{%
  pdftitle={\plaintitle}, pdfauthor={\plainauthor},
  pdfkeywords={\plainkeywords}, }

\begin{document}

\maketitle

\RaggedRight{} 

\begin{abstract}

We present an analytic provenance data repository that can be used to study human analysis activity, thought processes, and software interaction with visual analysis tools during exploratory data analysis.
We conducted a series of user studies involving exploratory data analysis scenario with textual and cyber security data.
Interactions logs, think-alouds, videos and all coded data in this study are available online for research purposes.
Analysis sessions are segmented in multiple sub-task steps based on user think-alouds, video and audios captured during the studies. 
These analytic provenance datasets can be used for research involving tools and techniques for analyzing interaction logs and analysis history.
By providing high-quality coded data along with interaction logs, it is possible to compare algorithmic data processing techniques to the ground-truth records of analysis history.
\end{abstract}



\newpage
\section{Introduction}
\label{sec:Introduction}
\import{\sectiondir}{1-introduction.tex}

\begin{figure*}[!t]
 \centering 
 \includegraphics[width=6.8in]{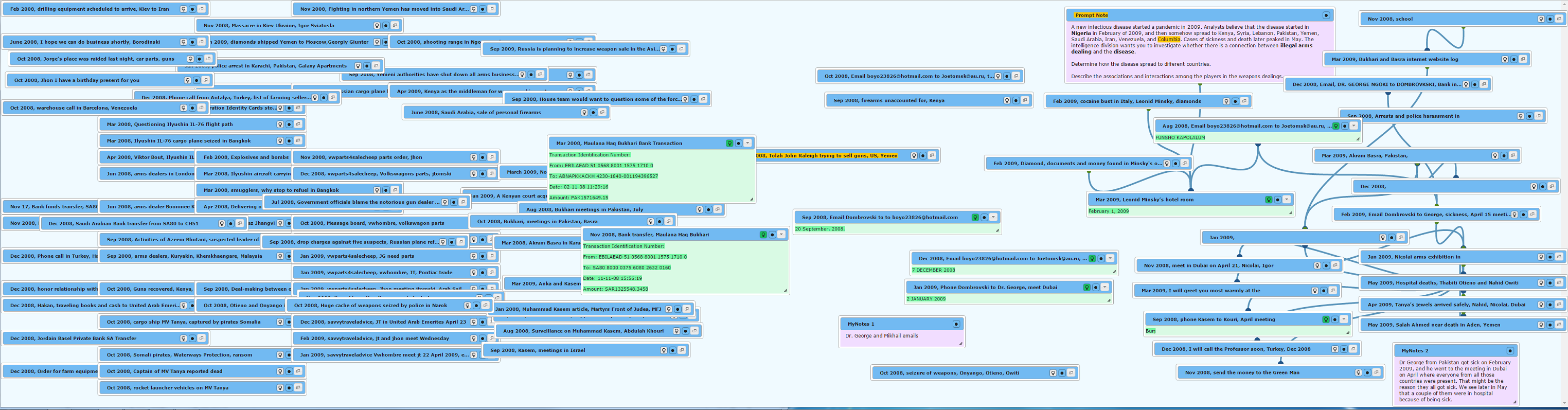}
 \caption{Screenshot of the document analysis tool used for collecting provenance data in the text analysis scenarios.
 All text documents are listed in a collapsed format with a random order on the left monitor at the beginning of the study.
 Documents have titles and users are able to drag and displace documents in the explorer tool space.
 In this tool user can write notes, move and link documents, highlight text, and search for keywords.
 }
 \label{fig:Exptool}
\end{figure*}

\section{Text Analysis Provenance Dataset}
\label{sec:Forward-Prediction}
\import{\sectiondir}{3-procedure.tex}


\section{Data Coding}
\label{sec:Results}
\import{\sectiondir}{4-thinkaloud.tex}

\section{Online Dataset}
\label{sec:Conclusion}
\import{\sectiondir}{5-online.tex}

\section{Acknowledgements}
This material is based on work supported by NSF 1565725.

\bibliographystyle{SIGCHI-Reference-Format}
\bibliography{ref}

\balance{} 

\end{document}

%% file: files/1-introduction.tex
Visual analytic tools assist analysts with exploratory inspection of large amounts of data to identify, understand, and connect pieces of information.
At a meta level, understanding analysis processes is important for improving tools, communicating analysis strategies, and explaining the evidence.
\textit{Provenance} for data analysis tracks the history of the analysis, including the progression of findings, interactions, data inspection, and visual state~\cite{ragan2016characterizing,mohseni2017threadsPoster}.
Analyzing user interactions and data provenance reveals more information about analysis process, helps in understanding how the user discovers insights, and is essential for understanding analysis behavior during open-ended data exploration tasks.

Designing visualization designs and techniques to study analysis processes requires sample analysis records for research and development.
Thus, our work contributes multiple analytic provenance datasets captured from user studies with high quality capture of participant interaction logs, think-aloud comments, screen capture, and transcribed notes from qualitative coding of sample analysis sessions from multiple data analysis scenarios.
To collect the provenance records, we conducted a set of user studies using basic visual data analysis tools appropriate for each scenario but generalizable enough to have similarities to many commonly used visualization software.
The datasets are fully anonymized and records are transcribed to for easy use by researchers interested in studying human data analysis behaviors.
Captured videos, user interaction logs and insight codings for all studies are available online ~\footnote{https://research.arch.tamu.edu/analytic-provenance/} for research purposes.

Currently, our provenance data repository contains records from two types of data analysis scenarios: textual intelligence analysis and multidimensional cybersecurity analysis scenarios.

%% file: files/3-procedure.tex
Our text analysis data is based on intelligence-analysis investigations from the publicly available VAST Challenge datasetsEach study session involved one of three intelligence analysis scenarios selected from the VAST Challenge data sets~\cite{scholtz2012reflection}, a set of synthetically created data sets and analysis scenarios designed to be similar to real-world cases and problems.
Specifically, our studies used data from the 2010 mini-challenge 1, 2011 mini-challenge 3, and 2014 mini-challenge 1.
All datasets contain various text records such as news articles, emails, telephone intercepts, bank transaction logs, and web blog posts.
For example, the 2014 data set involves articles about events and people related to missing individuals and violence related to a protest group in a fictional island.

Participants were tasked with gathering information and finding connections between events, people, places and times in the data sets.
Text documents varied in length from single sentences up to multiple paragraphs.
While all of the data was in plain-text format, some of the documents primarily consisted of numerical data related to financial transactions.
The 2010 data set had a total of 102 documents.
Due to the larger sizes of the 2011 and 2014 data sets, we reduced the number of included documents to 
accommodate constraints of user study duration for 90-minute sessions.
We used a subset of each data set to limit these two data scenarios to 152 documents.

All participants for each session were university students from varying majors, and ages ranged from 20 to 30. 
None of the users were expert in analytic tasks.
Participants used the document explorer tool and our cyber analysis tool desktop computer with two 27-inch monitors to analyze the data for 90 minutes.
At the beginning of study sessions, text explore tool and analysis task were explained to the users. 
Users had a 15 minutes time to work with the tool and ask questions prior to the start.
Currently, our provenance repository includes data records from 24 participants (6 female, 18 male) for the text analysis sessions.

To complete the analysis task, participants used a basic visual analysis tool (see Figure~\ref{fig:Exptool} on next page.
The tool supports spatial arrangement of articles, the ability to link documents, keyword searching, highlighting, and note-taking.
When loading the data in our document explorer tool, each document starts as collapsed with only its title visible.
Users could ``open'' any document by double-clicking the title bar or by clicking a dedicated button on the document's title bar, and this would expand the document to a window containing the text of the document.
The document could be collapsed back to the title in the same way.
Within an open document, users could highlight text by selecting it, right-clicking, and activating a menu item.
When a window has highlighted text, the window could be ``reduced to highlight'', which would hide all text in the document except for the highlighted content.
At the beginning of the study, documents were arranged in the left screen without a specific order or grouping.
Users clicked and dragged documents, freely re-arranging documents in the workspace.
They could also create editable notes windows in the same workspace.
When using the \textit{search} functionality, both matching words within windows and the windows themselves were highlighted.
Users could also draw connection lines across document windows, which created a line to denote relationships visually.

All user interactions at a rudimentary level like mouse movements and clicks are captured during the study using the text explorer.
Later we transform basic data log recorded from explorer tool to nine type of user actions, see Table~\ref{tab:table1}.
We associate analytical reasoning with different interactions available to the user, and later use it to modify the topic models.

Based on prior observations (e.g.,~\cite{MouseRef,goecks2000learning,jayarathna2015unified})  that mouse input can correspond with informational attentional. 
We use hovering the mouse over new document titles as users intend to explore new information.
Hovering mouse over document text shows reading interaction of the articles.

\begin{table}[tb]
  \centering
  \begin{tabular}{r l}
     \toprule
    \cmidrule(r){1-2}
    {\small\textbf{Interactions}}
    & {\small \textbf{Purpose}} \\
    \midrule
    Open documents & Explore new articles \\
    Read documents & Explore new information \\
    Search & Keyword search \\ 
    Highlight & Highlight document text \\
    Bookmark & Select documents \\
    Connect & Linking documents and notes\\
    Move documents & Arrange documents in screen \\
    Brush titles & Review document titles \\
    Creating Notes & Making sticky notes \\
    Writing Notes & Writing notes 

  \end{tabular}
  \caption{Types of interactions logged from the text analysis tool during the user studies.
  }~\label{tab:table1}
\end{table}

\begin{figure}[tb]
 \centering 
 \includegraphics[width=1.0\columnwidth]{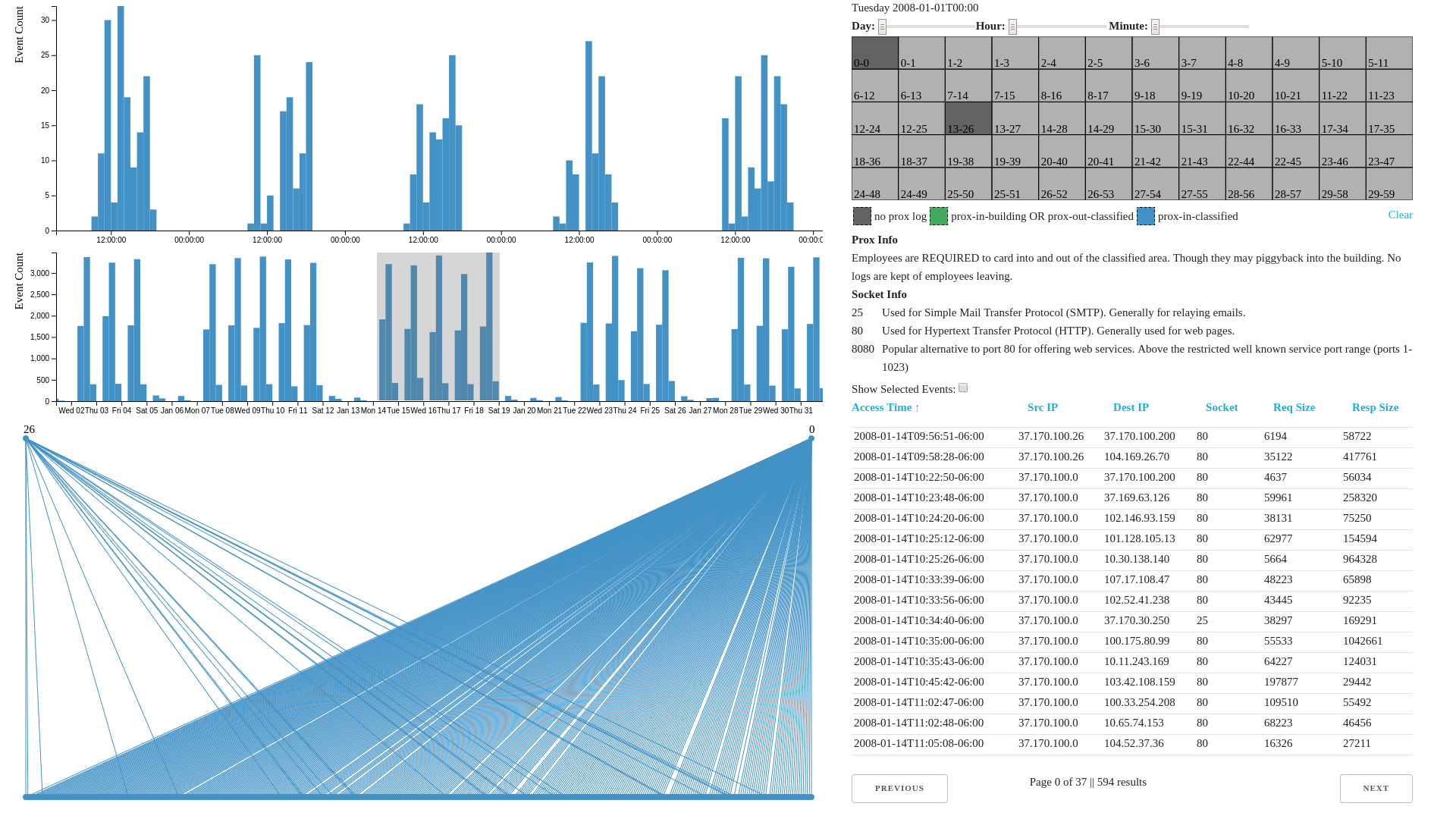}
 \caption{Screenshot of the cyber analysis tool used for collecting provenance data in the multidimensional analysis scenario. 
 Charts on the top left sides are detailed histogram and overview histogram, respectively. 
 A network graph is shown in the bottom left. 
 Boxes on the top right corner indicates the office view with slider tools.
 Below the office view is an information panel.
 Finally, at the bottom right, an IP traffic table shows detailed network data.  }
 \label{fig:Cybertool}
\end{figure}

\section{Multidimensional Data Analysis Dataset}

The multidimensional data analysis scenario currently has provenance records from 10 participants.
The analysis scenario used cyber analysis dataset was taken from 2009 VAST Challenge~\cite{grinstein2009vast}, mini-challenge 1.
The backstory of the scenario involves an employee of a fictional embassy trying to exfiltrate sensitive information to an outside criminal organization using office computers.
Participants explored this tabular multidimensional data set with a visual analysis tool comprised of multiple coordinated views.
Views include histograms of the traffic data, network graph, table of IP traffic details, and a work station layout showing proximity card status. 
Participants were asked to find the suspicious IP addresses used to transfer data to the criminal organization by exploring different views of the tool. 

Participants explored the multidimensional dataset to determine the suspicious behavior.
A static view of the tool is shown in Figure ~\ref{fig:Cybertool}.
The tool is divided into following 6 different views described in~\ref{tab:cybertable}.

Due to the large amount of data, filtering is required to find specific patterns in the data.
Brushing and linking is enabled in the overview histogram view.
Any portion of it can be selected and then the detailed histogram will change according to the selection.
Office view shows the current status of the employees inside embassy in that specific time frame using different color codes.
A slider tool also allows the participant to select specific time and day to check the employee status and network traffic.
Moreover, participants can select specific IP addresses from a multi-paged IP table for future reference.
User interactions with the tool is recorded in the form of mouse interaction and eye area of interest (AOI). 
Mouse tracking is done within the tool while eye tracking is performed using a Tobii EyeX, a standard eye tracking device that tracks the eye gaze fixation points.

\begin{table}[tb]
  \centering
  \begin{tabular}{r l}
     \toprule
    \cmidrule(r){1-2}
    {\small\textbf{Eye Area of Interest (AOI)}}
    & {\small \textbf{Mouse interaction Data logs}} \\
    \midrule
    Overview histogram & Brush start and end \\
    Detailed histogram & Mouse enter and bar click \\
    Network graph & Mouse enter \\ 
    Office view & Mouse click and slider move\\
    Information box & Mouse hover \\
    IP table & Page change and row select \\
  \end{tabular}
  \caption{Types of eye and mouse interactions logged from the cyber analysis tool during the user studies.
  }~\label{tab:cybertable}
\end{table}

%% file: files/4-thinkaloud.tex
In order for the provenance datasets to be useful for a wide range of research purposes, we prioritized the capture of users' thought processes and actions throughout the analysis activities.
We used a think-aloud protocol to capture participant's thoughts and insights during the study.
We transcribed users' think-aloud comments by watching the screen-captured videos of each session along with notes from the research team about observations from the study sessions.
Transcripts include all user's actions, talks, and time stamps of events.

\subsection{Coding for Text Analysis Dataset}

Two members of the research team reviewed all analysis records and identified times where user changed topic of the investigation.
We save all topic changes moments during the exploratory task and code them as topic changing (inflection) points.
For example, participant \textit{P7} was working on the third dataset and said ``I'm looking for these caterers at the executive breakfast'' and searches for ``caterers''.
The participant continues reading documents from this search for about 10 minutes.  
Then the user says ``I'm trying to figure out what the government was doing at the company'', which is a change in topic of investigation.
The user looks through titles and picks a couple of documents about government for about 8 minutes.
While reading new documents, \textit{P7 }finds out about the name ``Edward'' and is searching for incidents related to this name for the next 4 minutes.
There are also moments that user is done with current topic and wants to change the subject.
For instance, participant \textit{P3} working on the second dataset says ``Let's search for some keywords'' after 3 minutes of thinking and taking no actions. 
Then the user searches for keyword ``thread'' to find new articles about it.
Also, in many cases, topic changing does not include think-alouds, like opening a random document and continuing with that, or writing a note about an old topic, or even returning to an old topic after a while.

\subsection{Coding for Cyber Analysis Datset}

A similar approach was used to identify the inflection points in the cyber analysis data.
The research team identified key points by examining the task video and the audio of the think-aloud process.
Heuristics of marking the inflection points relied on the change in strategy attempted by the participant to complete the task.
Change in strategy can also be identified as the use of different views, different focal attributes within a view, or other means based on observations or verbal comments from the participant.

For example, participant \textit{Cyber-F} used the overview histogram to select some random times and tried to find unusual traffic pattern in the detailed histogram.
After spending about 5 minutes, the participant moved on to a new strategy involving the office view.
\textit{Cyber-F} then started using slider tool to find the proximity card status of different employees to know their current position and cross check with the IP table.
Another participant, \textit{Cyber-J} started the analysis by selecting each IP addresses and trying to find unusual traffic pattern in them. 
But with the large amount of data in the IP traffice table, the participant moved on to a new strategy after about 7 minutes.
The new strategy for \textit{Cyber-J} involved looking at the network graph to find unique destination IP addresses with large traffic. 
These changes in strategy are noted as inflection points by the coders and included in the transcripts.

%% file: files/5-online.tex
This analytic provenance datasets can be used for research involving tools and techniques for analyzing  interaction logs and analysis history.
By providing high-quality coded data along with interaction logs, it is possible to compare algorithmic data processing techniques to the ground-truth records of analysis history.
The Provenance Analytics Dataset is free and publicly available for research purposes. 
Captured videos, user interaction logs, the analysis tools used in the studies, and transcripts from think-aloud comments and observations from all studies are available online at \url{https://research.arch.tamu.edu/analytic-provenance/}.



 